\documentstyle[aps,multicol,epsfig]{revtex}
\begin{document}
\draft
\title{Polymer induced phase coexistence in systems of lamellar phases}
\author{Richard P. Sear$^{\dag}$}
\address{FOM Institute for Atomic and Molecular Physics,
Kruislaan 407, NL-1098 SJ Amsterdam, The Netherlands\\
{\rm email: sear@amolf.nl}}
\address{$^{\dag}$Address from October 1997:
Department of Chemistry and Biochemistry, The University of
California, Los Angeles, 405 Hilgard Avenue, Los Angeles,
California 90095-1569}
\date{\today}
\maketitle

\begin{abstract}
The effect of adding nonadsorbing polymer to
a lamellar phase of surfactant bilayers is studied theoretically.
We find that the
polymer produces
coexistence between two lamellar phases of different
layer spacings.
The coexistence region is a closed loop, as in the experiments
of Ficheux {\it et al.} [J. de Physique II {\bf 5} 823 (1995)].
Within our model the coexistence is driven by depletion.
\end{abstract}


%\newpage
\begin{multicols}{2}
\narrowtext
\section{introduction}

Many surfactants in solution form bilayers, i.e., two back-to-back
layers of surfactants molecules with their alkane-like tails in the centre
and only their hydrophilic headgroups in contact with the surrounding
water \cite{israelachvili92}.
These bilayers have been extensively studied both because of their
diverse behaviour \cite{gompper94} and because they
serve as simplified model systems for cell membranes
\cite{leibler89}.
In solution, bilayers
form a stack of parallel layers: a lamellar phase.
Here, we study theoretically the effect of adding non-adsorbing polymer
to an unbound lamellar phase \cite{ligoure93,nallet94,ficheux95}.
In particular, we attempt to understand some recent experimental results
\cite{ligoure93,ficheux95}.
A schematic of the system is shown in Fig. 1.
An unbound lamellar phase is one in which the
interactions between the bilayers are purely repulsive.
The polymer--bilayer interaction is also purely repulsive, leading
to a depletion in the density of polymer segments near the surface
of a bilayer \cite{daoud77,joanny79,gast83}. This
depletion has an associated free energy cost which is reduced
if two bilayers approach each other, then the two depletion layers
overlap \cite{gast83,vrij76,lekkerkerker95}.
Thus the polymer induces an attraction between the bilayers.
The attraction leads to phase separation: two lamellar phases with
different spacings between the bilayers coexist.

\section{Model and theory}

The lamellar plus polymer phase is described starting from the free
energy of a pure lamellar phase which is well known, see Refs.
\cite{israelachvili92,rand81,roux88,parsegian91}. Then semidilute
polymer solutions are discussed and the confining effect of the bilayers
on a semidilute polymer solution is estimated.
The confinement free energy of the polymer solution between the bilayers
is then added to the free energy due to the direct interaction between
the bilayers to produce a final free energy. This free energy is used
to calculate the phase behaviour of the bilayers plus polymer system.

\subsection{The interactions between the bilayers}

We treat the lamellar phase as a stack of
uniformly-spaced rigid plates; thus we can define a separation $d$ between
consecutive plates in the stack. This is the spacing between the
{\it surfaces} of
consecutive plates; the smectic period is $d+d_0$, where $d_0$ is the
(assumed constant) thickness of a bilayer. Of course, the bilayers
are not completely rigid but they bend on length scales
that are much larger than
$d$ or the size of a polymer coil \cite{gompper94}.
We neglect any interactions between
nonadjacent plates.

In the absence of polymer the free energy $F$ of a stack of $n$ layers, each of
area $A$, is taken to be \cite{israelachvili92,rand81,roux88,parsegian91}
\begin{equation}
\frac{F}{nA}=\frac{A_e}{d}+A_h\exp(-d/\lambda_h),
\label{freenp}
\end{equation}
The first of the two terms of
Eq. (\ref{freenp}) is the contribution of electrostatic
interactions and is long ranged while the second term is due to the
so-called hydration forces \cite{israelachvili92,rand81,parsegian91}.
The hydration interaction is
much shorter ranged than the electrostatic interaction, its range is of the
order of 1nm. We will not derive
Eq. (\ref{freenp}) here, it is a quite standard expression for the interaction
of a pair of charged bilayers separated by water.
The electrostatic term is derived from the
Poisson-Boltzmann equation for two charged plates separated by a gas of the
counterions \cite{israelachvili92,roux88}; without
added salt to screen the electrostatic interactions
between the bilayers. Moderate screening has been found
not to change the phase behaviour we predict.
$A_e$ is a positive constant that
is proportional to the surface charge density on the plates.
The second term in Eq. (\ref{freenp}) is essentially an empirical
description of the strong repulsion between two bilayers less than about
1nm apart. This repulsion is, to a good approximation, exponential.
Although it is strong,
$A_h\gg A_e/d$ for $d=1$nm, this repulsion is much shorter ranged than
the electrostatic interaction; $\lambda_h$ is small
$\simeq0.2$--$0.35$nm \cite{rand81,parsegian91}.
Equation (\ref{freenp}) does not include the contribution to the interlayer
interaction of the Helfrich or undulation forces
\cite{leibler89,helfrich78}, because these are dominated
by the electrostatic repulsions \cite{roux88} if, as here, the electrostatic
repulsions are not screened.

\subsection{The polymer--bilayer interactions}

As the polymer is nonadsorbing the bilayer--polymer
interaction is assumed to be a simple excluded volume interaction, i.e., each
bilayer excludes polymer segments from a planar volume.
The polymer will be in the semidilute regime,
we therefore use a simple scaling description of the structure and
thermodynamics of the polymer solution \cite{degennes79}.
The size of an isolated polymer coil of $N$ segments each of length $a$,
is measured by its radius of gyration $R_G$ which is given by
$R_G\sim aN^{3/5}$, in a good solvent  \cite{degennes79}.
The concentration of the polymer is specified by its volume fraction $\phi_b$.
The subscript $b$ is to remind us that
it is the volume fraction of the polymer far from the surface of a
bilayer. Near the surface of a bilayer the polymer's density drops
below $\phi_b$.
The average volume fraction of polymer in the
surfactant plus polymer system is not equal to $\phi_b$; it will be less.
The correlations (fluctuations) in
an isolated polymer coil extend across the entire coil; the correlation
length $\xi$ is thus $\xi\sim R_G$.
In a bulk solution of polymer, when the polymer coils
begin to overlap, interactions between different chains reduce the
correlation length $\xi$ \cite{degennes79}. The overlap
volume fraction $\phi_b^*$ is
$\phi_b^*=N^{-4/5}$.
This volume fraction is the boundary between the dilute and semidilute
regimes; below $\phi_b^*$ the interactions between chains can be treated
as a perturbation but above $\phi_r^*$ the interaction between a pair of chains
is $\gg k_BT$, where $k_B$ is Boltzmann's constant and $T$ is the
temperature. So, $\xi=R_G$ in the dilute regime, and
\begin{equation}
\xi\sim a\phi_b^{-3/4}~~~~~~~\mbox{in the semidilute regime}.
\label{xi}
\end{equation}
We require the form of the osmotic pressure of a
polymer solution, it is
\begin{equation}
\frac{\Pi a^3}{k_BT} \sim \left(\frac{a}{\xi}\right)^3\sim
\phi_b^{9/4}~~~~~~~\mbox{in the semidilute regime}.
\label{pp}
\end{equation}
The Helmholtz free energy of a polymer solution $F_p$
is also required. We can obtain the density dependent part
of the free energy, which is all that is required for determining phase
coexistence, from the thermodynamic relation $\Pi=-\partial F_p/\partial V$,
where $V$ is the volume of the solution. The volume derivative may
be transformed into a density derivative. Then the integration
of $\Pi$ as a function of density leads to
\begin{equation}
\frac{F_p a^3}{k_BTV}\sim{\rm const.}  + \phi_b^{9/4}.
\label{fp}
\end{equation}
Finally, the polymer's chemical potential $\mu_p$ is obtained from
$\mu_p=F_pa^3/(V\phi_r)+\Pi a^3/\phi$
\begin{equation}
\frac{\mu_p}{k_BT}\sim{\rm const.} + \phi_b^{5/4}.
\label{mup}
\end{equation}
All of the above is for a bulk polymer solution, i.e., a solution just
of polymer. In the presence of bilayers the solution is no longer uniform,
the polymer density at a point will depend on the distance from the
bilayers above and below the point.
However, at distances $\gg\xi$ from any bilayer, the polymer density
will reach a constant value.

We model the bilayer--polymer interaction as a hard wall--polymer
interaction \cite{daoud77}; there is no adsorption.
The correlation length $\xi$ is a measure of the distance over which
correlations decay. Therefore, if we introduce a single
bilayer, modelled
as a hard sheet, into a polymer solution it perturbs the solution
above and below its surface out to a distance of order $\xi$.
This is a much longer ranged perturbation than in a simple liquid
where the density of the fluid returns to its bulk value in a distance
of a few times $a$.
Within
$\xi$ of the surface the conformational freedom of the polymer chains
is reduced by the presence of the surface, reducing their
entropy and hence their density near the surfaces of the bilayers
\cite{daoud77,joanny79,gast83}.
The introduction of the bilayer has changed the free energy of the polymer
solution by introducing two polymer-wall `interfaces' (they are not true
interfaces as they do not separate two coexisting bulk phases).
This layer of reduced density is often called a depletion layer
\cite{lekkerkerker95}.
We require
the difference $\gamma$,
between the free energy of unit area of the layer
which is depleted in polymer and
the free energy of unit area of a layer of the same width
of the bulk fluid.
It is, essentially, the free
energy of a hard wall--polymer interface.
The picture is of a layer of thickness $\xi$ in which the density
drops from its bulk value $\phi_b$ to a much lower value.
The excess free energy of the depletion layer is then
approximately equal to the amount of work done in squeezing out the polymer
from a slab of unit area and width $\xi$ \cite{degennes79}.
Thus,
\begin{eqnarray}
\gamma\sim \Pi \xi.
\label{gamma}
\end{eqnarray}
This is the surface tension of one wall--polymer solution interface.
In a system of bilayers and polymer we can use this expression
provided both $\Pi$ and $\xi$ correspond to those values in a bulk solution
at the density of the polymer solution far, $\gg\xi$, from the
surface of any bilayer.
This can perhaps be made more clear if we consider that the
surfactant plus polymer system
is in equilibrium with a pure polymer solution across a
semipermeable membrane permeable to the polymer but not to the surfactant.
Then the volume fraction of the pure polymer solution in equilibrium with
our bilayers plus polymer solution, is the volume fraction at
which the $\Pi$ and $\xi$ of Eq. (\ref{gamma}) should be evaluated.

Each bilayer creates two
polymer depletion layers.
If the bilayers are sufficiently far apart,
$d\gg \xi$, then these layers are independent and
we have $2n$ depletion layers,
each of which contributes an amount to the free energy
given by Eq. (\ref{gamma}). However, as the bilayer separation $d$ is
reduced to $\sim \xi$ the
depletion layers of adjacent bilayers overlap.
When the layers overlap, the volume occupied
by these layers is less and therefore
the total free energy associated with these layers decreases.
It is this reduction in free energy which, we suggest, is responsible for
creating coexistence between two bilayer phases with different bilayer
spacings.

A difficulty now arises: the analysis which led to our estimate of
$\gamma$ in Eq. (\ref{gamma}) was purely qualitative,
whereas we now require a definite expression for how the
free energy of two depletion layers
changes as they coalesce.
The definite
expression is required to calculate a theoretical phase diagram in
order to compare with experimental phase behaviour.
The free energy of a pair of depletion layers $\gamma_2(d)$,
must be $2\gamma$ if there
are far apart, $d\gg \xi$. Within the scaling approach there is only
one relevant length scale: $\xi$. Thus, $\gamma_2$ is of the form
\begin{equation}
\gamma_2(d)=2\Pi\xi f(d/\xi);
\label{g2}
\end{equation}
that is, we are assuming that the density dependence is only through $\xi$.
The functional form of $f(d/\xi)$ is the same at all densities.
When $d\ll\xi$ the polymer density between the two
bilayers will be much less than in the bulk at the same chemical potential.
But if the polymer density between the plates is negligible then the
pressure exerted by the polymer inside the plates is negligible and
the pressure due to the polymer solution is the negative of the
pressure of the
bulk polymer solution, independent of the separation $d$. If the polymer
contribution to the pressure is a constant then its contribution to the
free energy must be $\Pi d$,
which forces $f(d/\xi)=d/(2\xi)$, for $d\ll\xi$.
In the other limit, $d\gg\xi$ $\gamma_2=2\gamma$ and so $f(d/\xi)=1$.
A function which satisfies both these requirements is
$f(z)= \mbox{tanh}(z/2)$.
This is merely an ad hoc function which is consistent with the little we
definitely know about the free energy of a polymer
solution in a slit \cite{joanny79,shih90}.
Brooks and Cates \cite{brooks93} have taken a different approach
to the same problem.
They obtain a free energy expression valid for all values of $d$ by forcing
the different scaling theory free energies \cite{daoud77}
in the different scaling regimes to match.
Polymer induced coexistence occurs for $d/\xi\sim1$
and so it is for these values of $d/\xi$ that $f(z)$ is required.
However, scaling theory is unable to predict the form of the free energy
when the parameter $d/\xi$ is of order unity.
So, where we find
coexistence, there is no reason to expect the free energy expression
of Brooks and Cates to be any more accurate
than the much simpler interpolation formula given here.
They found coexisting lamellar phases, as we will do, but they
did not find reentrance.
In order to find reentrance it is necessary to explicitly include
the contribution of hydration forces, which Brooks and Cates did not do.

For the bilayer plus polymer solution system
we use the variables $d$ and $\mu_p$, as these are most convenient.
A thermodynamic potential $\Gamma/(nA)$ for a pair of adjacent
interacting bilayers
is then obtained by adding the Helmholtz
free energy due to the repulsions between the bilayers $F/(nA)$,
to the grand potential of the polymer between the bilayers,
where the grand potential
includes an interfacial term, $\gamma_2$. We treat the slit between
a pair of bilayers as a thermodynamic phase.
The grand potential of the polymer in the slit
is then, by definition \cite{hansen86}, equal to $-\Pi Ad +\gamma_2 A$,
as the volume of the phase is $Ad$.
As we are fixing the chemical potential not the
density of the polymer the bulk part of the contribution of the polymer is not
its Helmholtz free energy but $-\Pi V$.
So, from Eqs. (\ref{freenp}) and (\ref{g2})
\begin{equation}
\frac{\Gamma}{nA}=\frac{F}{nA} - \Pi d +  2\Pi\xi f(d/\xi)
\label{gammap}
\end{equation}
The right hand side of Eq. (\ref{gammap}) is $\Gamma$ per slit.
The three terms are the direct repulsion between the two bilayers,
the bulk part of the grand potential of the `phase' between the slits
and the interfacial part of the grand potential of this `phase'.
Equation (\ref{gammap}) is correct in the
two limits of pure bilayers and pure polymer, and the third term is
a crude approximation to the additional cost of the
polymer--bilayer depletion layer.
In the derivation of Eq. (\ref{gammap}) we have
implicitly treated the bilayers plus polymer mixture not as a mixture
but as a polymer solution between a stack of hard walls
Because of this assumption $\Pi$ is
unaltered by phase separation and is the same in the coexisting phases.
It is not the pressure of the bilayer plus polymer system.

Coexistence can be found from the common tangent condition. This
leads to the requirement that
at coexistence the two phases have equal values of
$\partial [\Gamma/(nA)]/\partial d$ and
$\Gamma/(nA)-d(\partial [\Gamma/(nA)]/\partial d)$.
These two equalities determine the values of $d$ in the two coexisting phases.

\section{Comparison with experiment}

Ficheux {\it et al.} \cite{ficheux95} studied a lamellar phase
of the surfactant sodium dodecylsulphate (SDS) in
an aqueous solution of the polymer poly(ethyleneglycol) (PEG).
We set the parameters of $\Gamma$ (\ref{gammap}) to be reasonable
for their system. There are 5 parameters:
$A_e$, $A_h$, $\lambda_h$, $a$ and $R_G$.
For the purposes of calculation we express them in reduced units. These
units are defined using the length of a polymer segment $a$ as our length
scale and as our energy scale we use
$k_BT$.
Then $A_e$, $A_h$, $\lambda_h$ and $R_G$ are expressed in units of
$k_BT/a$, $k_BT/a^2$, $a$ and $a$, respectively.
We take $k_BT=4\times10^{-21}$J and $a=0.3$nm; the first
is approximately its value at room temperature and the second is reasonable
for a flexible polymer.
We set
$A_e=0.5k_BT/a$, $A_h=50k_BT/a^2$, $\lambda_h=a$ and $R_G=10a$.
These values correspond to
$A_e=6.67\times10^{-12}$Jm$^{-1}$, $A_h=2.22$Jm$^{-2}$,
$\lambda=0.3$nm and $R_G=3$nm.
Ficheux {\it et al.} report an $R_G=2.9$nm \cite{ficheux95}.
A hydration decay length of 0.3nm is approximately what has been found
in a number of experiments \cite{israelachvili92,rand81,parsegian91}.
Our value of $A_h$ is comparable to the estimate given by
Roux and Safinya of 2Jm$^{-2}$ \cite{roux88} and our
$A_e$ is comparable to the $4.49\times10^{-12}$Jm$^{-1}$ obtained from
(the dominant term in) the solution of
the Poisson-Boltzmann equation \cite{roux88}.

Using this set of parameters, we have calculated the phase diagram of
a stack of bilayers in the presence of a polymer solution;
the $\Pi$--$d$ plane of the diagram is shown in Fig. 2.
There is a {\it closed} region of phase coexistence, as in the
experiment of Ficheux {\it et al}; it
is bounded at high and low polymer pressures by two
critical points. Note, however, that Ficheux {\it et al} do not
rule out the possibility of the polymer adsorbing onto the bilayers.
If too little or too much polymer is added no phase coexistence is found.
The presence of a closed loop of phase separation is distinctly unusual.
We will show why theory predicts it in the hope that
our theory provides a reasonable description of the
experiment of Ficheux {\it et al.}.
It is worth noticing that the closed loop is entirely within the
semidilute regime. The dashed curve in Fig. 2 shows $2\xi$ which
only reaches its $2R_G$, indicating the crossover to the dilute regime,
at values of $\Pi$ below those of the phase coexistence loop.

In the absence of polymer, the free energy (\ref{freenp}), and the pressure,
of a stack of bilayers is a monotonically decreasing function of $d$.
The plates repel each other at all separations due to the repulsive
electrostatic and hydration interactions.
In the presence of polymer,
there is an additional term in the thermodynamic potential (\ref{gammap}),
from the depletion layers.
It is the overlap of these
layers that is driving phase separation.
For a stack with
$d$ a few times $\xi$,
the depletion layers contribute nearly
$2\gamma$ per bilayer. However, if this phase separates into two phases:
one with a large $d$ and one with a $d\lesssim 2\xi$,
then the total free energy of the polymer decreases because of the reduced
total volume of the solution within $\xi$ of a bilayer surface.
Thus the total contribution of the depletion layers
to the free energy is reduced, at a cost in free energy due to
the bilayer--bilayer interactions. As the parts of the free energy 
from bilayer--bilayer interactions increase more rapidly than linearly,
the phase separation necessarily increases this part of the free
energy.

The total free energy available from overlapping the polymer
depletion layers is of the order
of $\Pi\xi$. Thus, phase separation will only occur if this is sufficient
to overcome the repulsion of the bilayers. Of course, this repulsion
decreases as $d$ increases; so if $\xi$ is large the polymer
depletion layers overlap when the bilayers are far apart and repel each other
only weakly. Conversely, 
if $\xi$ is small the repulsions are strong.
The polymer solution can only overcome the bilayer repulsions
and produce two bilayer phases if both $\Pi$ and $\xi$ are sufficiently large.
In the semidilute regime $\xi$ decreases as $\Pi$ increases and so
the two requirements are, to an extent, antagonistic. They are only
satisfied for a range of $\Pi$.

The pressures due to the electrostatic and hydration
interactions are $A_e/d^2$ and $(A_h/\lambda_h)\exp(-d/\lambda_h)$,
respectively. In order for the polymer to control the phase behaviour
the osmotic pressure of the polymer must be greater than both of these,
at the separations at which the polymer depletion layers
begin to overlap. So, we have two inequalities which must be satisfied
in order for there to be phase separation,
\begin{equation}
\Pi > A_e/\xi^2.
\label{electro}
\end{equation}
and
\begin{equation}
\Pi>\frac{A_h}{\lambda_h}e^{-\xi/\lambda_h}.
\label{hyd}
\end{equation}
Both of these inequalities only give the correct order of magnitudes.
The correlation length $\xi$ varies with $\Pi$ as $\xi=a(\Pi/T)^{-1/3}$.
Using this relation we see that
the right hand side of Eq. (\ref{electro}) varies
as $\Pi^{2/3}$, more slowly than the left hand side.
Thus Eq. (\ref{electro}) provides a lower bound to the polymer osmotic
pressure required to induce phase coexistence.
For values of $\Pi$ above this bound the overlap of the interfaces releases
enough free energy to overcome the electrostatic repulsions.
Now, consider Eq. (\ref{hyd}).
If we express $\xi$ in Eq. (\ref{hyd}) in terms of $\Pi$ we see that
the right hand side of Eq. (\ref{hyd}) increases with $\Pi$ as
$\exp(-\Pi^{-1/3})$. This increase is more rapid than linear and so
Eq. (\ref{hyd}) provides an upper bound to the
polymer osmotic pressure required to induce phase coexistence.
The presence of these lower and upper bounds ---
due to the long and short ranged parts
of the bilayer repulsions, respectively ---
leads straightforwardly to a closed loop
coexistence region.
There is slight subtlety in that the loop is $d\approx15$ which
is $5\lambda_h$. The hydration repulsion is able to have an effect
at such long range because the electrostatic repulsions and polymer
induced attractions are delicately balanced in the coexistence region.
Analogous behaviour has been predicted by Gast {\it et al.} \cite{gast83}
when polymer is added to a suspension of charged spherical particles.
There, as here, the decreasing range of the polymer induced attraction
can cause reentrant miscibility.
When the spheres repel each
other with a long ranged electrostatic repulsion then the range
of the polymer induced attraction between the spheres must
have a longer ranged than this repulsion

As polymer is added the dielectric
constant of the solution between the bilayers
decreases \cite{zaslavsky89}; decreasing the
repulsive electrostatic interactions.
Our neglect of this effect may cause us to underestimate the size of the
coexistence loop but incorporating a polymer concentration dependent
$A_e$ will not change the qualitative phase behaviour.

\section{Conclusion}

Our theoretical model yields a closed-loop phase-coexistence
region in the phase diagram. The model is a simple, additive, combination
of a standard model for the interaction between charged bilayers and
a crude scaling form for the free energy of a polymer solution in a slit.
It provides a description of the behaviour of a mixture of
surfactant bilayers and semidilute nonadsorbing
polymer on a length scale of 1--10nm;
the length scale on which the entropy of the polymer solution and the
repulsive bilayer--bilayer interactions are competing.
Thus, we
provide an explanation of the phase behaviour in terms of behaviour
at mesoscopic length scales.
The model shows qualitatively the same behaviour as found in experiment
\cite{ficheux95}, without any fitting. This encourages us to believe that
our theory describes correctly the mechanism behind the experimental
observations.

It is a pleasure to acknowledge D. Frenkel for illuminating discussions and
J. Polson for a careful reading of the manuscript.
I would like to thank The Royal Society for the award of a fellowship
and the FOM institute AMOLF for its hospitality.
The work of the FOM Institute is part of the research program of FOM
and is made possible by financial support from the
Netherlands Organisation for Scientific Research (NWO).

%\newpage

\end{multicols}
\widetext

\newpage
\begin{figure}
\begin{center}
\epsfig{file=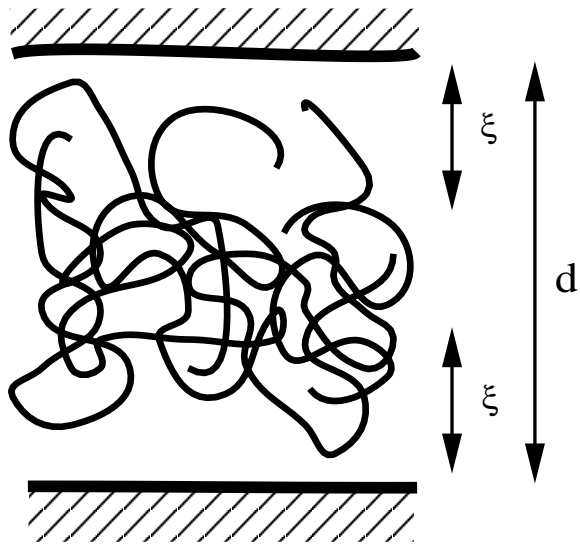, width=3.0in}
\vspace*{1.0in}
\end{center}
\caption{
Schematic of a pair of bilayers with a polymer solution
between them.
The thick curves denote the surfaces of the bilayers and the polymer
is represented by the entangled thinner curves.
$d$ is the spacing between the bilayers and $\xi$ is the polymer
correlation length.
Note the low density of
polymer near the surfaces of the bilayers: the depletion layer.
}
\end{figure}

\newpage
\begin{figure}
\begin{center}
\epsfig{file=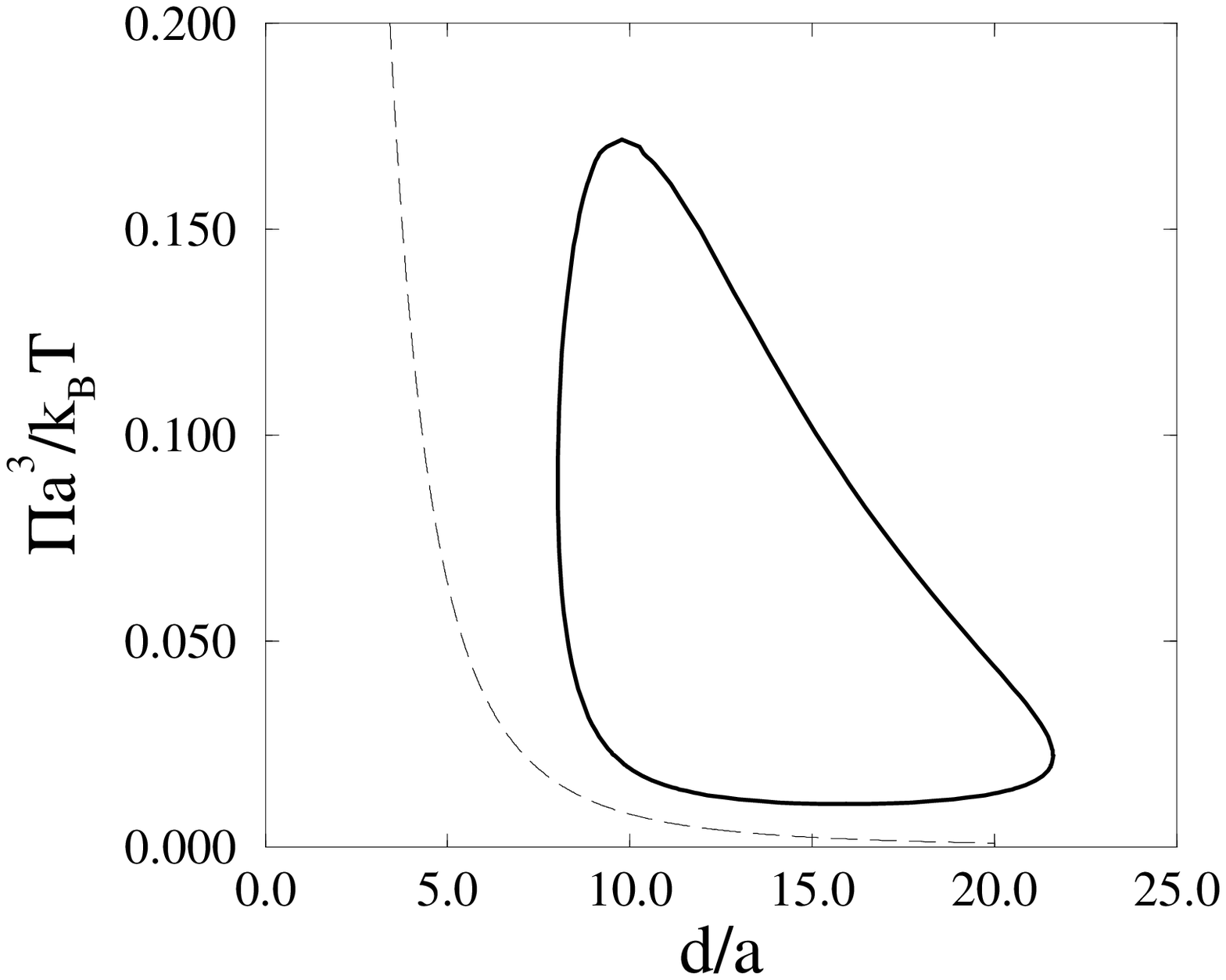, width=5.0in}
\end{center}
\vspace*{-1.2in}
\caption{
The phase diagram of a bilayer plus polymer solution
system, as calculated from the thermodynamic potential (\ref{gammap}), with
the parameter values given in the text.
$\Pi$ is the osmotic pressure of the polymer
and $d$ is the spacing between the surfactant bilayers.
The solid curve separates the one and two-phase regions and the dashed curve
is $2\xi$ as a function of the osmotic pressure of the polymer $\Pi$.
As $\Pi$ decreases $\xi$ increases until it reaches the radius
of gyration, equal to $20d$ here, whereupon it remains constant.
The thin horizontal lines are tie lines.
}
\end{figure}


\begin{thebibliography}{99}

\bibitem{israelachvili92} J. N. Israelachvili,
{\it Intermolecular and Surface Forces}
(Academic Press, London, 1992).

\bibitem{gompper94} G. Gompper and M. Schick,
{\it Self-Assembling Amphiphilic Systems}
(Academic Press, London, 1994).

\bibitem{leibler89} S. Leibler,
{\it Statistical Mechanics of Membranes and Surfaces},
D. Nelson, T. Piran and S. Weinberg eds. (World Scientific, Singapore, 1989)

\bibitem{ligoure93} C. Ligoure, G. Bouglet and G. Porte,
{\it Phys. Rev. Lett.} {\bf 71} (1993) 3600.

\bibitem{nallet94} F. Nallet, D. Roux, C. Quilliet, P. Fabre and
S. T. Milner,
{\it J. de Physique II} {\bf 4} (1994) 1477.

\bibitem{ficheux95} M. F. Ficheux, A. M. Bellocq, F. Nallet,
{\it J. de Physique II} {\bf 5} (1995) 823.

\bibitem{daoud77} M. Daoud and P. G. de Gennes,
{\it J. de Physique} {\bf 38} (1977) 85.

\bibitem{joanny79} J. F. Joanny, L. Leibler and P. G. de Gennes,
{\it J. Polymer Sci.: Polymer Phys. Ed.} {\bf 17} (1979) 1073.

\bibitem{gast83} A. P. Gast, C. K. Hall and W. B. Russell,
{\it Faraday Discuss. Chem. Soc.} {\bf 76}, 189 (1983).

\bibitem{vrij76} A. Vrij,
{\it Pure Appl. Chem.} {\bf 48}, 471 (1976).

\bibitem{lekkerkerker95} H. N. W. Lekkerkerker, P. Buining, J. Buitenhuis,
G. J. Vroege and A. Stroobants,
{\it Observation, Prediction and Simulation of Phase Transitions in
Complex Fluids}, M. Baus {\it et al.} eds. (Kluwer, Dordrecht, 1995).

\bibitem{rand81} R. P. Rand,
{\it Ann. Rev. Biophys. Bioeng.} {\bf 10} (1981) 277.

\bibitem{roux88} D. Roux and C. R. Safinya,
{\it J. de Physique} {\bf 49} (1988) 307.

\bibitem{parsegian91} V. A. Parsegian, R. P. Rand and N. L. Fuller,
{\it J. Phys Chem.} {\bf 95} (1991) 4777.

\bibitem{helfrich78} W. Helfrich,
{\it Z. Naturfosch.} {\bf 33a} (1978) 305.

\bibitem{degennes79} P. G. de Gennes,
{\it Scaling Concepts in Polymer Physics}
(Cornell, Ithaca, 1979).

\bibitem{shih90} W. Y. Shih, W. H. Shih and I. A. Aksay,
{\it Macromolecules} {\bf 23} (1990) 3291.

\bibitem{brooks93} J. T. Brooks and M. E. Cates,
{\it J. Chem. Phys.} {\bf 99} (1993) 5467.

\bibitem{hansen86} J. P. Hansen and I. R. McDonald,
{\it Theory of Simple Liquids} (Academic Press, London, 2nd edn, 1986).

\bibitem{zaslavsky89} B. Yu. Zaslavsky {\it et al},
{\it J. Chem. Soc. Faraday Trans. I} {\bf 85} (1989) 2857.


\end{thebibliography}
\end{document}